\documentclass[journal,twocolumn,twoside]{IEEEtran}
\usepackage{epsf,xspace,relsize}
\usepackage[rflt]{floatflt}
\usepackage[final]{graphics}
\usepackage{epsfig,color,colortbl}
\usepackage{epstopdf}

\begin{document}
\title{A Cherenkov Radiation Detector with High Density Aerogels}
\author{
	Lucien~Cremaldi~\IEEEmembership{Member,~IEEE,}
         David A. Sanders~\IEEEmembership{Member,~IEEE,}
         Peter Sonnek,
         Don J. Summers,
         Jim Reidy, Jr. 
\thanks{Lucien Cremaldi, David A. Sanders, Peter Sonnek, and Donald J. Summers are with the Department of Physics and Astronomy, University of Mississippi, University, MS, 38677 USA \hfill\break 
(email: cremaldi@phy.olemiss.edu, tel: +1-662-915-5311)
}%
\thanks{Jim Reidy, Jr. is with Oxford High School, Oxford, MS, 38655 USA}
}

\markboth{IEEE Transactions on Nuclear Science~Vol.~x, No.~x, June~2009}%
{Cremaldi \MakeLowercase{\textit{et al.}}: A Cherenkov Radiation Detector with High Density Aerogels}

\maketitle 
  
\begin{abstract}
We have designed a threshold Cherenkov detector at the Rutherford-Appleton Laboratory to identify muons with momenta between 230 and 350 MeV/c.  We investigated the properties of three aerogels for the design. The nominal indexes of refraction were n = 1.03 , 1.07, 1.12, respectively.  Two of the samples are of high density aerogel not commonly used for Cherenkov light detection. We present results of an examination of some optical properties of the aerogel samples and present basic test beam results.
\end{abstract}



\section{Introduction}

\IEEEPARstart{C}{ooled} muons are required for a Neutrino Factory based on a muon storage ring \cite{Ayres}, \cite{Albright}, \cite{Berg} and for a muon collider \cite{Neuffer}, \cite{Ankenbrandt}, \cite{Summers}, \cite{Palmer}.  The Muon Ionization Cooling Experiment (MICE) \cite{Drumm}, \cite{Yoshida} at Rutherford-Appleton Lab is the first test of the ionization cooling concept for muon beams. To establish muon cooling a high purity muon beam is positively identified in the momentum range 230 to 350 MeV/c by both time-of-flight and Cherenkov techniques. Silica aerogel has been successfully used in Cherenkov particle identification detectors in the past \cite{Cantin},  \cite{TASSO}, \cite{BELLE}, \cite{trans}. The Spherical Neutral Detector (SND) also uses high density silica aerogel \cite{Barnyakov}.  Based on 
 these and our work
 with gas threshold Cherenkov counters \cite{Bartlett}, we chose high density silica aerogels from the Matsushita Electric Works \cite{Matsushita} for its high quality and hydrophobicity (water repellant nature). 

We selected aerogels with indexes of refraction $n$=1.07 ($p_{th}^{\mu}$=278 MeV/c ) and $n$=1.12 ($p_{th}^{\mu}$=210 MeV/c), these aerogels nicely filling the gap between gas and liquid radiators for our momentum range. These  high density aerogels  scatter heavily in the UV and visible and it is not well known if they can be effectively used for Cherenkov radiators. In our work we compared the scattering and absorptive properties of our high density samples to a commonly used silica aerogel with index $n$=1.03. The panel sizes were nominally 115~mm x 115~mm x 11.5~mm. We report on index of refraction, transmission, and scattering measurements, and we measure the light collection yields with particle beams.  The panels are displayed below in Fig.~1.

\begin{figure}[ht]
\centerline{
\includegraphics[width=80mm]{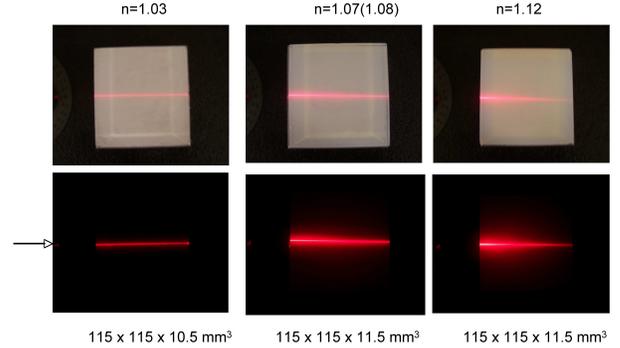}}
\caption{Photograph of the aerogel panels under study with incident laser light. The beam is incident from the left. }
\label{aerogels}
\end{figure}

\section{Index of Refraction Measurements}
The index of refraction of the aerogel samples was measured in a straight-forward method using the total internal reflection technique described below.  A red laser diode (638~nm) 
was mounted on a rotating platform with angle scale shown in Fig.~2. We could pinpoint the angle of total internal reflection $r$  to a fraction of a degree.  The  angle $p$ was determined by swiveling the laser beam until the point of total internal reflection (angle $s$ = 90$^o$).  Using the relation     \begin{eqnarray} n = 1 + \sin^2(p) \end{eqnarray} we determined the index of refraction relating index $n$ to angle $p$. The method has the advantage  of yielding a small measurement error $ \Delta n \leq 0.005$ once the aerogel block is carefully aligned w.r.t. the laser beam. Due to slight irregularity at the aerogel edge we could measure the angle $p$ to within  $\pm 0.5^o$. The results are reported in Table I.

\begin{figure}[ht]
\centerline{
\includegraphics[width=80mm]{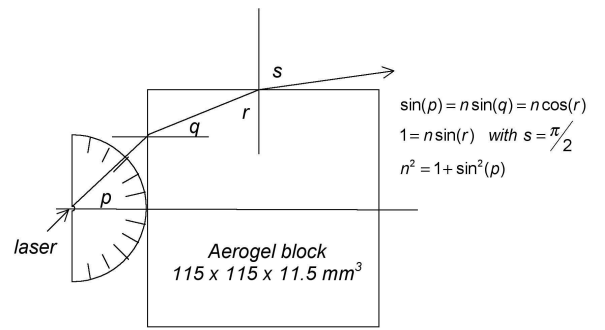}}
\caption{Schematic of index of refraction measurement by total internal reflection method.
Angle $s$ is adjusted to $\pi$/2 by the method.}
\label{fig_1}
\end{figure}

Our measurement for the $n$=1.03 aerogel (as certified by Matsushita \cite{Matsushita}) agreed well
at wavelength 638~nm.  The change in index (chromatic dispersion) for the 1.03 aerogel is reported \cite{hermes} to be $\Delta n = 0.0015 $ over the range (325-635)~nm, well within errors of our measurement technique. 

\begin{table}
\label{tb:one}
\caption{ Table I: Index of refraction measurements  with the internal refraction technique ($\lambda = 638~nm $)of the $n$=1.03,1.07,1.12 aerogel samples from Matsushita.}
\renewcommand{\arraystretch}{1.3}
\begin{tabular}{cccc}\hline 
Aerogel      & $p$ degrees & $n$ \\ \hline
Aerogel 1.03 (YI-30) & $14.0^o\pm 0.5^o$  &  $1.029\pm 0.002$ \\ \hline
Aerogel 1.07 (HY-80)& $22.2^o\pm 0.5^o$  &  $1.069\pm 0.003$ \\ \hline
Aerogel 1.12 (HY-120) & $29.1^o\pm 0.5^o$  &  $1.112\pm 0.004$\\ \hline
\end{tabular}
\end{table}

\section{Index of Refraction by Density}
It is known that the aerogel index of refraction tracks with density in a linear approximation
where coefficient $\alpha \approx$ 0.28- 0.30  for many aerogel samples
\begin{eqnarray} n = 1 + \alpha \times \rho ~. \end{eqnarray}

We carefully measured the dimensions of the aerogel samples with optical technique and determine they were within $\pm$ 500 $\mu$m of our 11.5 cm  specification to Matsushita. Some thickness variation over the surface was noticed of order 100-200 $\mu$m.  The sample masses were measured on a precision scale to find the aerogel density  $\rho = m/V$ in (g/cc). 
  
The indexes in Table II are determined from a Matsushita linear extrapolation $n$ = 1.000 + 0.279$\rho $ which works well for $\rho \leq 0.25$ g/cc.  For the higher density aerogel 112 sample our index measurement and density measurement may indicate some quadratic correction is needed, but still within error of the linear approximation. 

\begin{table}
\caption{Table II: Determination of the index of refraction with the density technique.}
\label{tab:two}
\renewcommand{\arraystretch}{1.3}
\begin{tabular}{cccccc}\hline 
Aerogel     & $V$(cc) & $m$(g) & $\rho$ (g/cc)& n=1.000 + 0.279$\rho~\cite{Matsushita}$ \\ \hline
$n$=~1.03 & 138.86  & 15.634   & 0.113$\pm0.003$   & $1.032\pm0.007$  \\ \hline
$n$=~1.07 & 152.08  & 38.792   & 0.255$\pm0.005$   & $1.071\pm 0.005$  \\ \hline
$n$=~1.12 & 152.08  & 56.293   & 0.370$\pm0.005$   & $1.103\pm 0.005$   \\ \hline
\end{tabular}
\end{table}

\section{Measurements of Forward Transmittance}
\begin{figure}[ht]
\centerline{
\includegraphics[width=70mm]{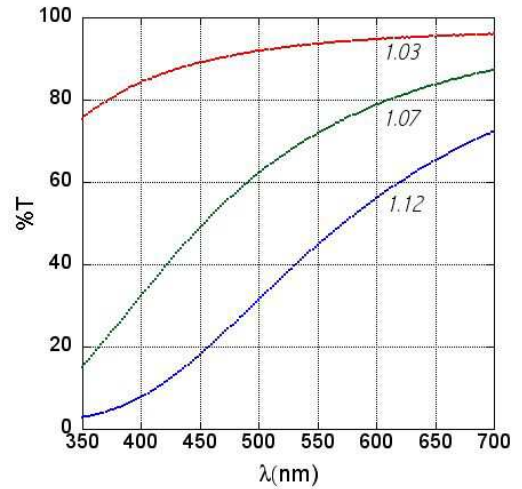}}
\caption{Measurement of aerogel percent transmissions versus wavelength for the three aerogel tiles. }
\label{fig_3}
\end{figure}

We performed a transmission $T$ scan  of the 11.5~mm thick aerogel tiles with a UV/VIS Spectrophotometer \cite{perkelmer} to extract the aerogel scattering coefficients.  The samples are positioned in the spectrometer 
so only the forward-going light cone ($\Delta\theta_{1/2} \leq 3^o$) is accepted  in to the spectrophotometer integrating sphere. We define this as a working definition of transmission $T$ and do not correct for reflection at the sample interfaces $\leq 0.7$\% or surface scattering effects $\leq 0.3$\%. This yields a systematic measuring error on $T$ of about $0.8$\%, well within our survey goal. The results are shown in Fig. \ref{fig_3}. 

If we assume Rayleigh scattering off small aerogel mesopores (2-50~nm)  and macropores ($d \approx$ 80$\pm$10~nm) the thin sample scattering intensity $I$  at distance $R$ and angle $\theta$ can be described by the well known formula \cite{trans}:

\begin{eqnarray}
  I = I_0\left({    {1+\cos^2(\theta)} \over{2R^2} }\right) \left({{2\pi}\over{\lambda}}\right)^4\left({{n^2-1}\over {n^2+2}}\right)^2               
({d \over 2})^6~,
\label{eq_rayleigh}
\end{eqnarray}
displaying the angular,  wavelength, and refractive index dependance. 
Following Eq. \ref{eq_rayleigh}, the compounded forward transmission $T$ through an aerogel of thickness $t$ can be parameterized by \cite{trans}:
\begin{eqnarray}
T = T_0~e^{-C~t/\lambda^4~.} 
\label{eq_trans}
\end{eqnarray}

\noindent{where $T_0$ is the wavelength independent transmittance, and $C$ is the Rayleigh scattering coefficient. We fit each transmission curves to Eq. \ref{eq_trans}.~ Fits to the $n$=1.12 aerogel are reported for $\lambda\geq$ 500~nm due to poor fit $\chi^2$'s below this value. The results are given in Table III. We also calculate the wavelength dependent Rayleigh transmission length $X_0=\lambda^4/C$ at 500~nm in Table III for comparison.} 

\begin{table}
\caption{Table III: Transmission fits to equation \ref{eq_trans} ($\lambda(nm)$).}
\label{tab:three}
\renewcommand{\arraystretch}{1.3}
\tabcolsep=1.8mm
\begin{tabular}{cccccc}\hline 
Aerogel     & $t$(cm) & $T_0$ \%  & C($10^{10}$)& $X_0$(500~nm) &$\chi^2$ \\ \hline
$n$=~1.03~(350-700 nm) & 1.05    & 97.56     & 0.36  & 17.4 cm & 1.5  \\ \hline
$n$=~1.07~(350-700 nm) & 1.15    & 98.05     & 2.47  & 2.5 cm  &2.7    \\ \hline
$n$=~1.12~(500-700 nm) & 1.15    & 96.98     & 6.16  & 1.0 cm   &0.7    \\ \hline
\end{tabular}
\end{table}

\section{Scattering Length}
\begin{figure}[ht]
\centerline{
\includegraphics[width=70mm]{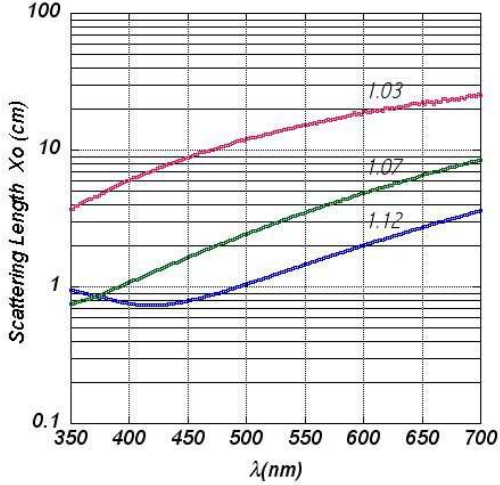}}
\caption{Measurement of the scattering length X$_0$ versus wavelength for the three aerogel samples.  }
\label{fig_4}
\end{figure}

Although Eq.(4) is a good parameterization of the data for the different aerogel samples, we take a second and more direct approach and define a scattering length $X_0(=\lambda^4/C)$  by $T = T_0 e^{-t/X_0}$. To extract $X_0$ we take the ratio of the transmission spectrum of a single tile $T_1$ to the transmission spectrum of two tiles $T_2$.  The ratio $T_2/T_1$ yields the aerogel scattering length $X_0=-t/\ln(T_2/T_1)$  as a function of wavelength. The results are plotted in Fig. \ref{fig_4}. The fit and ratio method of section IV. are in reasonable agreement.  

\section{Intensity versus Scattering Angle}
\begin{figure}[ht]
\centerline{
\includegraphics[width=50mm]{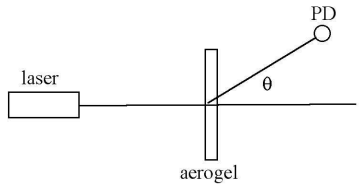}}
\caption{Apparatus sketch for measuring the in-plane light intensity vs. angle.  }
\label{fig_scatter}
\end{figure}

\begin{figure}[ht]
\centerline{
\includegraphics[width=70mm]{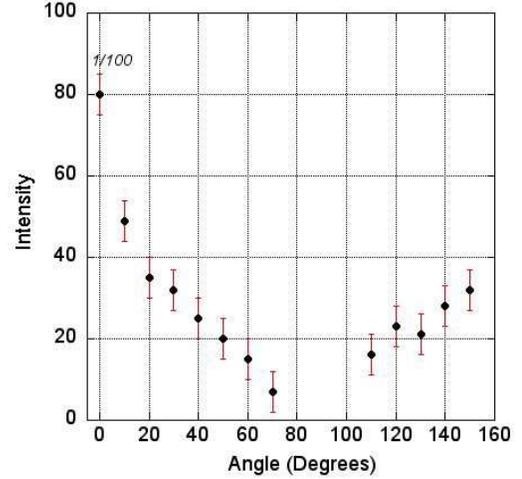}}
\caption{Intensity(unscaled) versus angle in the light plane for the $n$=1.12 aerogel tile. The measurement at $\theta$=0 is scaled by $1/100$.}
\label{fig_6}
\end{figure}

To investigate the angular scattering pattern of the aerogel, green light (532~nm) from a laser diode was directed with incidence normal to the face of an aerogel $n$=1.12 tile (11.5~mm thick), as depicted in Fig. \ref{fig_scatter}.
The light intensity  versus scattering angle $\theta$ was measured by a photodiode (PD) located at a distance of 10cm from the sample in the plane. The results are shown in Fig. \ref{fig_6} for the in-plane measurements. The photo detector subtended a solid angle of $\Delta\Omega$=0.08$\pi$ capturing about  2\% of the scattered light. The forward laser transmission was measured to be $I/Io$ = 32\%, indicating a scattering length of $X_0 \approx 1.3$cm, consistent with previous measurements at 532~nm (see Fig. \ref{fig_4}).

Although the measurement errors are large, one can see the light intensity is strongly peaked in the forward direction $\theta=0^0$, within a few degrees of the beam, as expected. 
The light scatter versus angle shows a strong forward-backward scatter as predicted by the Rayleigh law, suggesting that back-scattered Cherenkov light can be collected by a properly designed reflector. A small measurement asymmetry about $\theta$=90$^o$ was noticed suggesting misalignment or some other systematic in the measurement. The data was not corrected for optical path length difference versus angle. 
   
\section{Beam Test}
\begin{figure}[ht]
\centerline{
\includegraphics[width=90mm]{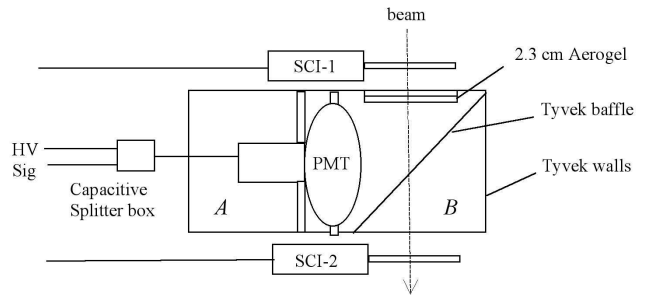}}
\caption{Apparatus for cosmic ray and beam test of aerogel panel.}
\label{tb_app}
\end{figure}

A small prototype Cherenkov detector was used to measure photoelectron response of the aerogel samples in cosmic rays (the results not given here) and a 120 GeV/c proton beam ($\beta=1$) at Fermilab.  The apparatus, depicted in Fig. \ref{tb_app}, consisted of  two chambers.  In chamber $A$ an 8 inch PMT \cite{electron_tubes} was placed with dome extending in to the second aerogel sample chamber $B$. In the sample chamber $B$ two aerogel tiles were mounted to the top face at the center of the compartment.  A 1/2 mil PVC window covered the aerogel, held it in place and protected it from contaminants.  The compartment $B$ was lined with a Tyvek \cite{tyvek} diffuse reflector. It was found that a simple angled Tyvek baffle increased the light collection by 10\% and was inserted. A 2-fold trigger was set up between two scintillator paddles SCI-1 and  SCI-2 which matched the aerogel's 11.5 x 11.5 cm$^2$ dimension. The data were collected with a LRS 2249 charge integrating ADC with a 60 ns gate (1/4\,pC/channel).  The PMT was run at a moderate gain of +1600V. The ADC was gated on the beam coincidence in the trigger counters. 

\begin{figure}[ht]
\centerline{
\includegraphics[width=80mm]{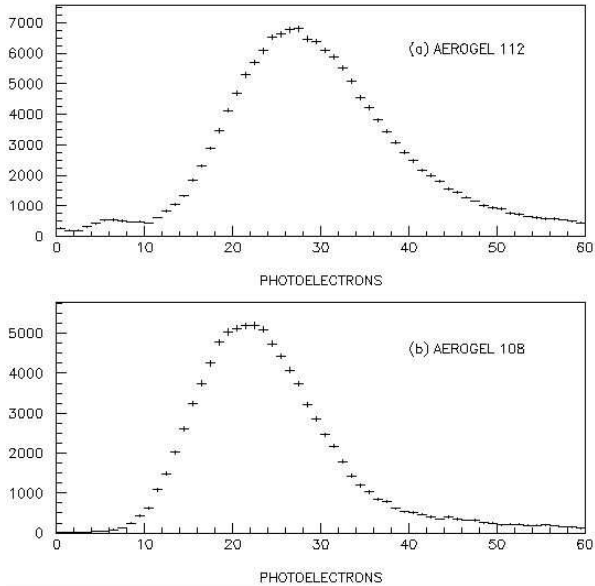}}
\caption{120 GeV/c proton beam exposure. $\beta=1$ photo-electron peaks correspond to $27\pm2$ and $22\pm2$ photoelectrons respectively for the 2.3cm thick aerogel (a) $n$=1.12 and (b) $n$=1.07 samples, respectively.}
\label{tb_data}
\end{figure}

Based on a predicted photo-electron (pe) yield of $N_{pe}\approx 90 (1-{{1}\over{n^2}})L\cdot\epsilon$ for vertical $\beta=1$  protons with 100\% collection efficiency $\epsilon$=1, a proton beam would generate 26, 39  photoelectrons in the aerogel $n$=~1.07 and $n$=~1.12 samples, respectively (see Table \ref{tab:four}). We measured 22$\pm$2 and 27$\pm$2,  with a $\pm$2 uncertainty from our gain calibration (see Fig. \ref{tb_data}). 

\begin{table}
\caption{Table IV: Beam test result}
\label{tab:four}
\renewcommand{\arraystretch}{1.3}
\begin{tabular}{ccccc}\hline 
Aerogel          & $L$(cm) & $N_{pe}^{pred}$ & $N_{pe}^{meas}\cdot \epsilon$ & $\epsilon$  \\ 
            &               &         ~$(\epsilon=1)$                   &                                 &             \\ \hline
$n$=~1.07   & 2.3     & 26.2                         & 22$\pm$2                      & $0.84\pm 0.08$     \\ \hline
$n$=~1.12   & 2.3     & 39.0                         & 27$\pm$2                      & $0.69 \pm0.05$     \\ \hline
\end{tabular}
\end{table}

The measurement results and estimated light collection efficiencies $\epsilon$ are listed in Table IV. In this configuration the efficiencies are about $70\pm 10$\%, typical for a diffusely reflecting collection box. This gives good evidence that the light absorption in the aerogel 1.12 and 1.07 panels is small. 

\section{Summary}

We have made some of the the first measurements on high density aerogel $n$=1.07 and $n$=1.12 samples, from Matsushita, to be used in a beamline Cherenkov application. The transmission and scattering properties were investigated and compared to standard $n$=1.03 aerogel. Little evidence for strong light absorption in the high density aerogels was found in our test beam exposure, indicating the possibility of further uses of the high density aerogels in particle detector and beamline applications. 

We want to thank Mac Yamauchi and Hiroshi Yokogawa at Matsushita Electric Works and the Panasonic Corporation for their help in preparation of the aerogel samples.  We also thank Koji Yoshimura at KEK-Tsukuba for his help in obtaining the samples.




\end{document}